\documentclass[5p,authoryear,final,times]{article}

\usepackage{graphicx} % Include figure files
\usepackage{txfonts}

\newcommand{\mc}[3]{\multicolumn{#1}{#2}{#3}}
\newcommand{\nuc}[2]{${}^{#1}$#2}

\begin{document}

\huge
\centerline{Activities of $\gamma$-ray emitting isotopes}
\centerline{in rainwater from Greater Sudbury, Canada}
\centerline{following the Fukushima incident}

\vskip0.5cm

\large
\centerline{B.\,T.~Cleveland, F.\,A.~Duncan, I.\,T.~Lawson,}
\centerline{N.\,J.\,T.~Smith, E.~V\'azquez-J\'auregui}

\vskip0.5cm

\large
\centerline{SNOLAB, 1039 Regional Road 24}
\centerline{Lively ON, P3Y 1N2, Canada}

\normalsize
\hyphenation{Fu-ku-shim-a Ma-no-lo-pou-lou Cle-men-za}

\begin{abstract}

We report the activity measured in rainwater samples collected in the
Greater Sudbury area of eastern Canada on 3, 16, 20, and
26~April~2011.  The samples were $\gamma$-ray counted in a germanium
detector and the isotopes \nuc{131}{I} and \nuc{137}{Cs}, produced by
the fission of \nuc{235}{U}, and \nuc{134}{Cs}, produced by neutron
capture on \nuc{133}{Cs}, were observed at elevated levels compared 
to a reference sample of ice-water. These elevated activities are 
ascribed to the accident at the Fukushima Dai-ichi nuclear reactor 
complex in Japan that followed the 11~March earthquake and tsunami.
The activity levels observed at no time presented
health concerns.

\end{abstract}

%\begin{keyword}
%Fukushima \sep 131I \sep 134Cs \sep 137Cs \sep Fission products
%\end{keyword}

\section{Introduction}

The nuclear accident in the Fukushima area in Japan released
radioisotopes to the atmosphere which have been measured in 
several locations in
Asia \cite{Bolsunovsky,Fushimi,Momoshima},
North America \cite{Bowyer,Leon,Norman,Sinclair,MacMullin},
and Europe \cite{Clemenza,Manolopoulou,Pittauerova},
as the radioactivity spread around the Earth.  We
report here the measurement of several isotopes in water samples
collected in eastern North America during April~2011, from 3 to
7~weeks after the Fukushima incident.

\section{Experimental Methods}

To investigate the dispersal of radioactivity from the Fukushima
incident we collected samples of rainwater in aluminum trays in
Greater Sudbury, Ontario on 3, 16, 20, and 26~April, the first rainy
days that followed 11~March.  A reference sample of ice water from
Meatbird Lake in Lively, Ontario was also collected on 3~April.
Within 1--2~d of their collection the water samples were passed
through Whatman Grade 1 filters (medium porosity, $>11~\mu$m) and
then poured into 1~L polyethylene Marinelli beakers.  Filtration was
necessary to remove particulate material that was present in the
samples because they were collected at ground level under windy
conditions.  The volume of all samples was very close to 1~L.  The
beakers were sealed, encapsulated in nearly air-tight bags, and
transported to the SNOLAB underground laboratory where they were
$\gamma$-ray counted by a high-purity germanium detector.  To minimize the
background from ambient \nuc{222}{Rn} in the mine air~\cite{hpge}, the samples
began to be counted immediately after their arrival underground.  The
duration of counting was 1~d except for the ice sample which was
counted for 2~d.  Data on the samples and counting periods are given in
Table~\ref{sample_data}.

\begin{table}[htb]
\centering
\begin{footnotesize}

\caption{Data on samples.  Dates of sample collection and start of
counting are given in day of year 2011 in Eastern Standard Time
(GMT-5~h).  Dead time during counting was negligible.}

\begin{tabular}{l l l l l}
\hline
              & Date      & Volume & Date counting & Counting \\
Sample        & collected & (mL)   & began         & time (d) \\
\hline
Ice 3 April   &   93.46   &  953   &  96.315       & 2.01     \\
Rain 3 April  &   93.96   &  857   &  95.345       & 0.93     \\
Rain 16 April &  106.48   &  935   & 109.275       & 1.02     \\
Rain 20 April &  110.57   & 1015   & 123.314       & 0.97     \\
Rain 26 April &  116.44   & 1050   & 124.356       & 1.0      \\
\hline
\end{tabular}
\label{sample_data}
\end{footnotesize}
\end{table}

% Uncorrected 131I rates            Decay factor
% Ice 3 April     17.8 +/- 2.9        0.782
% Rain 3 April    593 +/- 39          0.887
% Rain 16 April   50.3 +/- 7.0        0.786
% Rain 20 April   10.4 +/- 3.5        0.333
% Rain 26 April   1.2 +2.3 -1.2       0.505

The dimensions of the Ge detector crystal are 63-mm length by 67-mm
diameter and its efficiency for the 1333-kev $\gamma$-rays from a
\nuc{60}{Co} source is 47\% relative to a 3-inch by 3-inch NaI(Tl)
detector.  The FWHM resolution of the detector at 1333~keV is
1.9~keV.  To reduce local background the detector is shielded by
2~inches of high-purity copper and 8~inches of lead.  The detector
shield is enclosed in a sealed copper box through which pure nitrogen 
from liquid nitrogen boil-off is flowed at 2~L/min to purge \nuc{222}{Rn}.  
The efficiency of the detector for $\gamma$-rays has been measured 
with standard sources of known decay rate.

\begin{figure*}[hbtp]
\centering
\includegraphics[width=0.9\hsize]{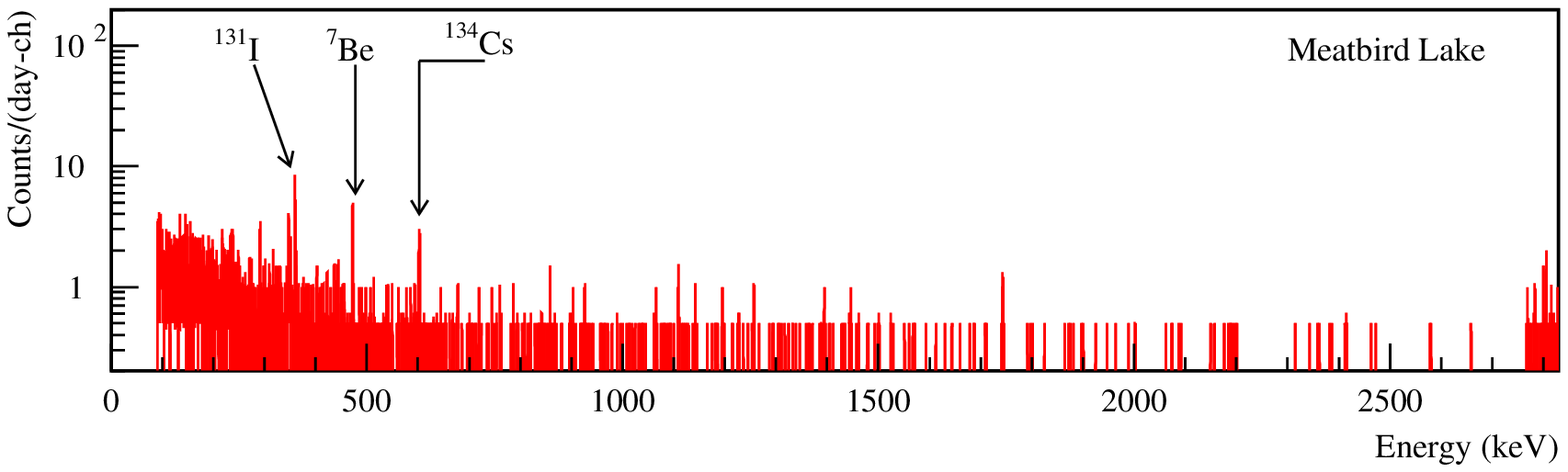}
\vskip-2.4cm
\includegraphics[width=0.9\hsize]{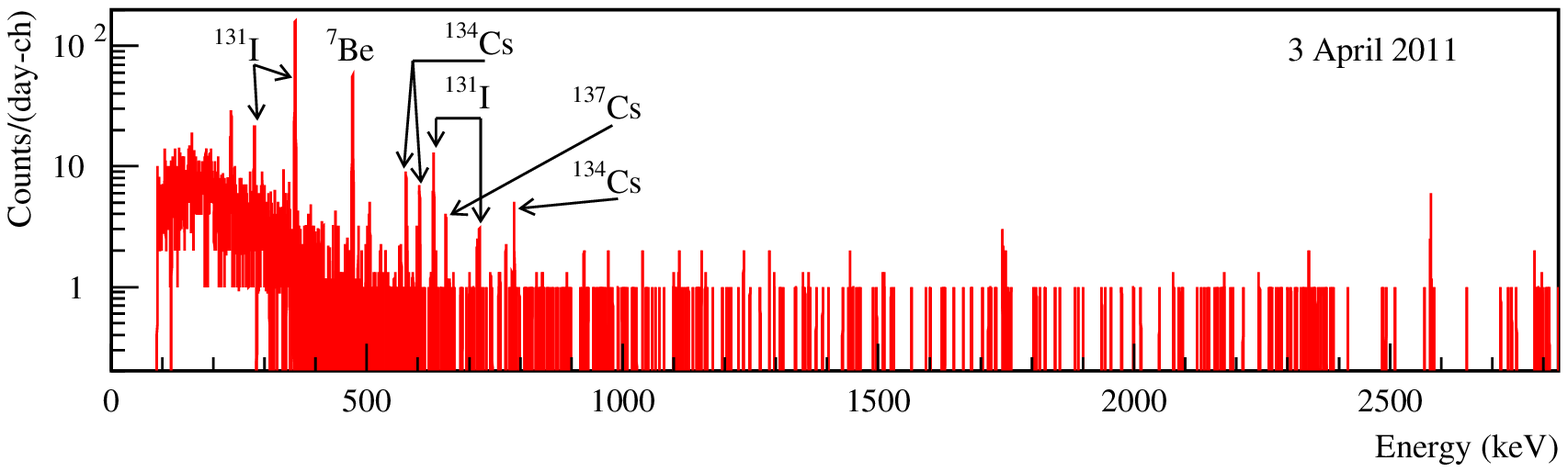}
\vskip-2.4cm
\includegraphics[width=0.9\hsize]{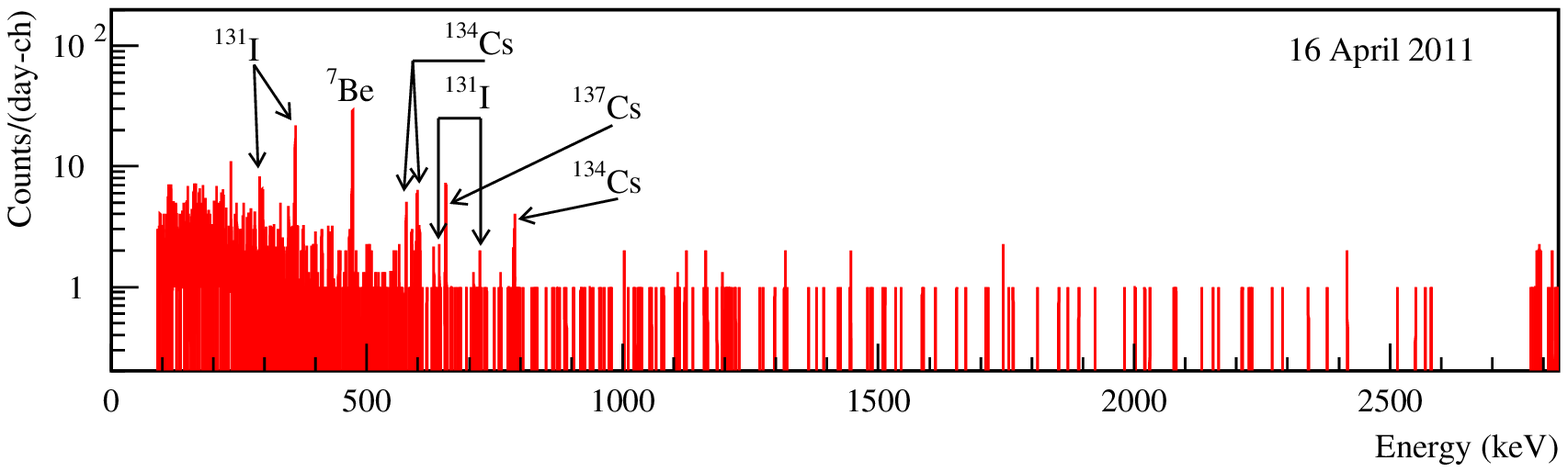}
\vskip-2.4cm
\includegraphics[width=0.9\hsize]{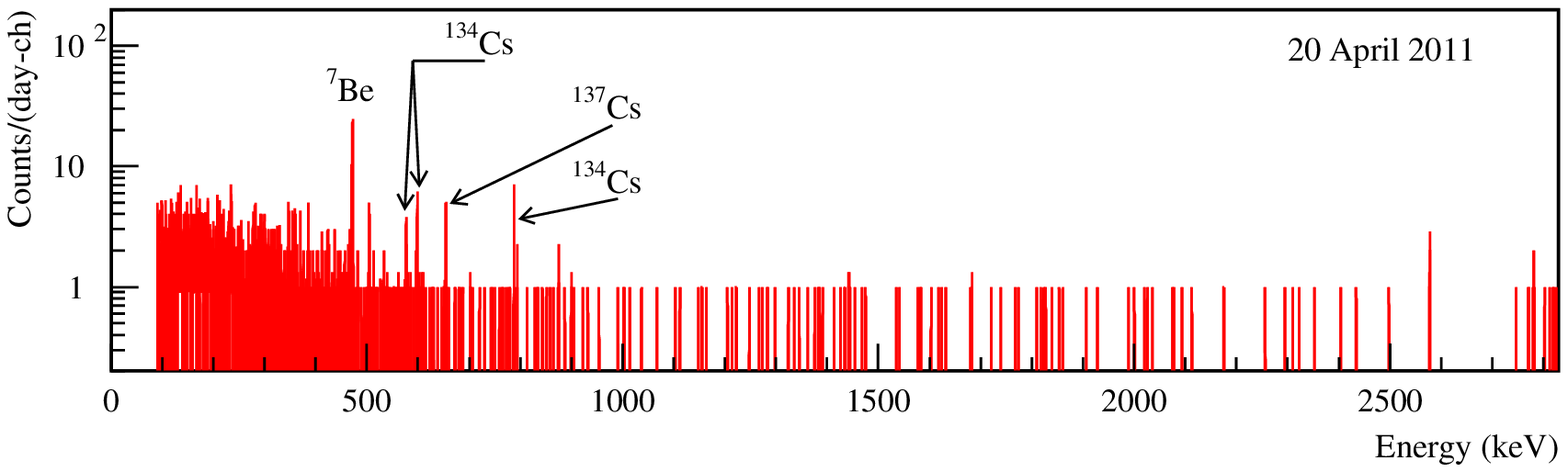}
\vskip-2.4cm
\includegraphics[width=0.9\hsize]{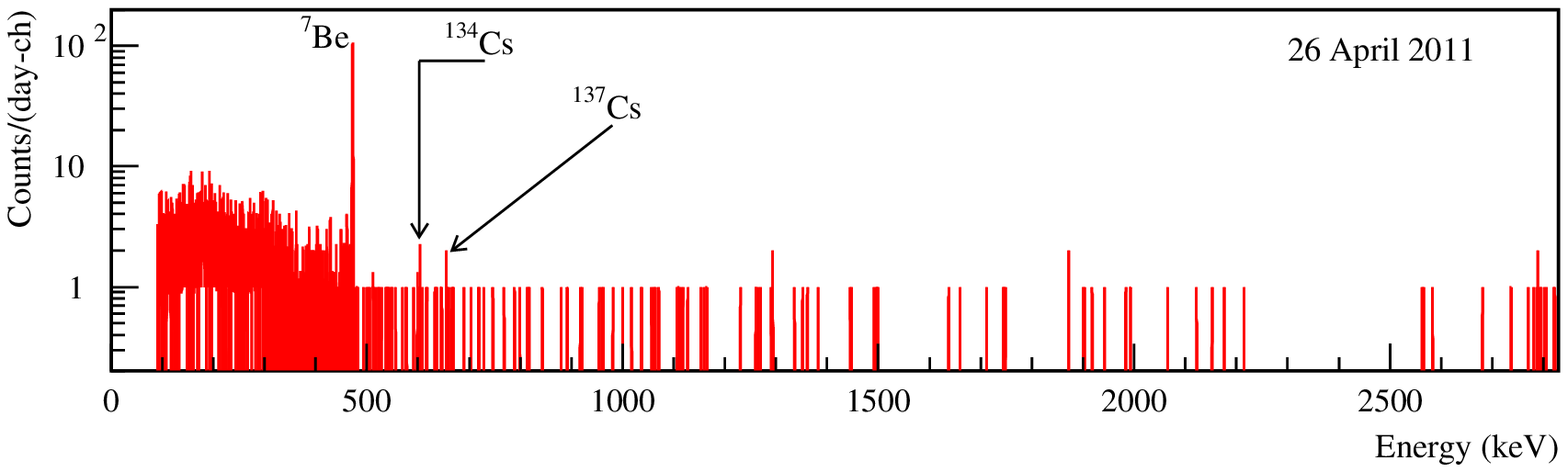}
\vskip-1.0cm

\caption{Energy spectra of ice water sample from Meatbird Lake and
rainwater samples.}

\label{spectra_samples}
\end{figure*}

\begin{table*}[hbtp]
\centering
\begin{footnotesize}

\caption{Specific activity of radionuclides detected in the ice water
and rainwater samples.  Uncertainty includes statistical and
systematic components and is given with 68\% confidence.}

\begin{tabular}{l l l l l l l}
\hline
              & \mc{6}{c}{Specific activity (mBq/kg)} \\ \cline{2-7}
Sample        & \nuc{137}{Cs}       & \nuc{134}{Cs}       & \nuc{131}{I}        & \nuc{238}{U} progeny & \nuc{232}{Th} progeny & \nuc{7}{Be} \rule{0pt}{2.5ex} \\
\hline
Ice   3 April & $<0.4$              & $0.4^{+0.8}_{-0.4}$ & $22.8\pm3.7$        & $26.1\pm3.1$ & $2.4\pm1.7$ & $80\pm18$ \\
Rain  3 April & $11.0\pm4.1$        & $8.3\pm2.4$         & $668\pm44$          & $32.8\pm6.2$ & $101\pm11$  & $1900\pm180$  \\
Rain 16 April & $22.7\pm4.9$        & $16.8\pm2.9$        & $64.0\pm8.9$        & $ 9.9\pm4.8$ & $9.4\pm2.9$ & $835 \pm95$   \\
Rain 20 April & $19.1\pm4.7$        & $13.1\pm2.5$        & $31\pm11$           & $<4.8$       & $5.9\pm2.7$ & $770 \pm90$   \\
Rain 26 April & $0.9^{+1.7}_{-0.9}$ & $0.7\pm0.7$         & $2.4^{+4.6}_{-2.4}$ & $3.3\pm2.6$  & $<0.3$      & $2700\pm240$  \\
\hline
\end{tabular}

\label{samples}
\end{footnotesize}
\end{table*}

This detector is usually used to measure the activity of samples of
materials that are being considered for use in one of the SNOLAB
experiments, all of which must be made from extremely low-background
components.  The detector sensitivity is 1~mBq/kg (0.1~ppb) for
\nuc{226}{Ra}, 1.5~mBq/kg (0.3~ppb) for \nuc{228}{Th}, and 21~mBq/kg
(0.7~ppm) for \nuc{40}{K}.  Further information on the detector and
its use is given in ~\cite{hpge}.

The samples were collected close to the SNOLAB laboratory which is
located at $46^\circ 28.5'$~N latitude, $81^\circ 12.0'$ W longitude.
The counting facility is 2092-m underground where the cosmic-ray muon
flux is $3.31 \times 10^{-10}$/(cm$^2$s)~\cite{sno}.

\section{Results and Interpretation}

The raw energy spectra from each sample are shown in
Figure~\ref{spectra_samples}.  Peaks are evident from the emission of
$\gamma$-rays by \nuc{137}{Cs} at 661.7~keV; by \nuc{134}{Cs} at 569.3,
604.7, and 795.9~kev; and by \nuc{131}{I} at 284.3, 364.5, 637.0, and
722.9~keV.  $\gamma$-ray lines with intensity greater than background are
also apparent at 477.6~keV from \nuc{7}{Be} (mainly produced by
cosmic-ray spallation on \nuc{14}{N} and \nuc{16}{O}) and at several
other energies from the decays of the progeny of \nuc{226}{Ra} and
\nuc{228}{Th}, which were present in the water samples as impurities.

\begin{figure}[hbtp]
\centering
\includegraphics[width=1.0\hsize]{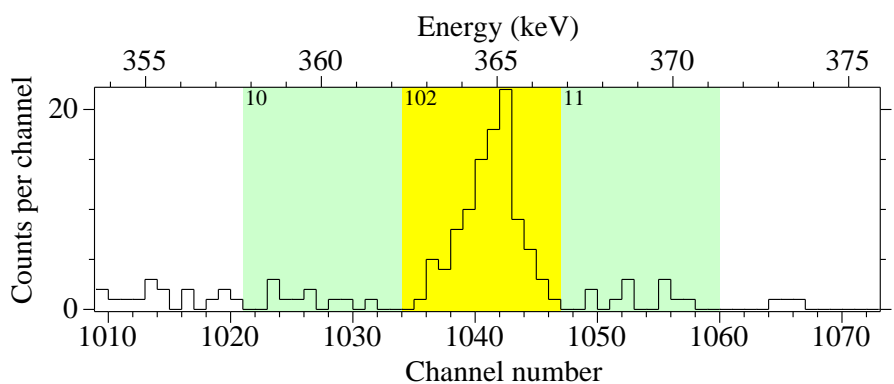}

\caption{Peak at 364.5~keV produced by \nuc{131}{I} decay in the
rainwater sample collected on 16~April and measured on 19 April.  The
counting time was 1.02~d and 102~events were recorded in the peak.
Background in equal energy-width windows was 10~events (below peak)
and 11~events (above peak).}

\label{example_spectrum}
\end{figure}

The region of a peak in a typical spectrum is shown in
Fig.~\ref{example_spectrum}.  The number of counts above background
in each peak is determined by counting the number of events in a
2-FWHM region centered at the peak and subtracting half the number of
events in regions of equal energy both above and below the peak.
\nuc{137}{Cs} is an exception as there also must be subtracted a
constant background in the peak of $1.72 \pm 0.13$ counts/d.  This
latter activity is contamination internal to the Ge detector crystal
housing.

The specific activities of the isotopes observed in the water samples
are given in Table~\ref{samples}.  Because of its short half-life
the \nuc{131}{I} activities have been corrected to the time of sample
collection.  For those isotopes that produced more than one peak we
checked that the activity inferred from each peak was in agreement
within uncertainty and we give their weighted average.

The observation of the short-lived isotope \nuc{131}{I} and the high
concentrations of \nuc{137}{Cs} and \nuc{134}{Cs} indicate a recent
release into the atmosphere of typical reactor-produced isotopes.

\begin{figure}[hbtp]
\centering
\includegraphics[width=1.0\hsize,viewport=29 29 502 292]{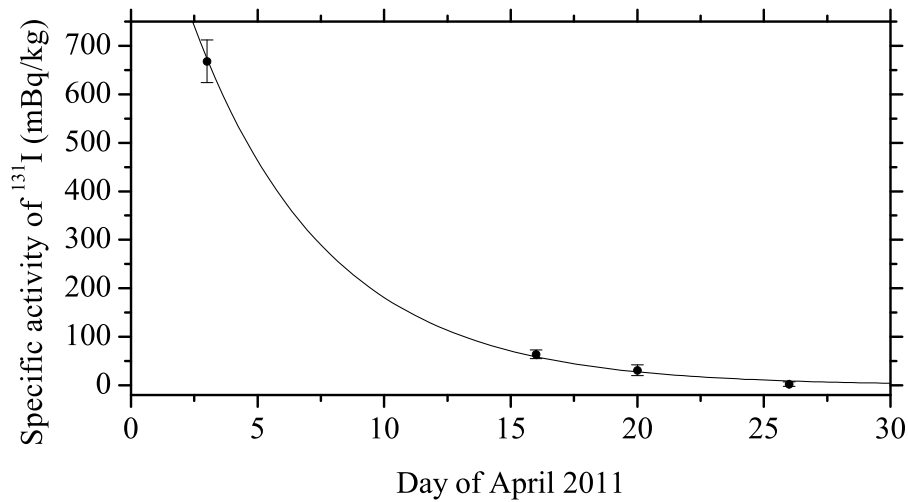}

\caption{Specific activity of \nuc{131}{I} vs time.  The solid line
is a fit of the specific activity $A$ as a function of time $t$ to
the function $A(t) = A(0) \exp(-t/t_1)$ where $A(0)$ and $t_1$ are
constants whose best fit values are $A(0) = 1183 \pm 99$~mBq/kg and
$t_1 = 5.33 \pm 0.30$~d.}

\label{131_vs_time}
\end{figure}

Figure~\ref{131_vs_time} shows the decay of activity of \nuc{131}{I}.
The half-life of the observed decay is 3.7~d, considerably less than
the 8.0~d half-life of \nuc{131}{I}. We presume this is due to the transport of
the radioactivity over our measuring location and washout of the
isotope from the atmosphere.

\vskip +5ex

\begin{figure}[hbtp]
\centering
\includegraphics[width=1.0\hsize,viewport=29 29 502 292]{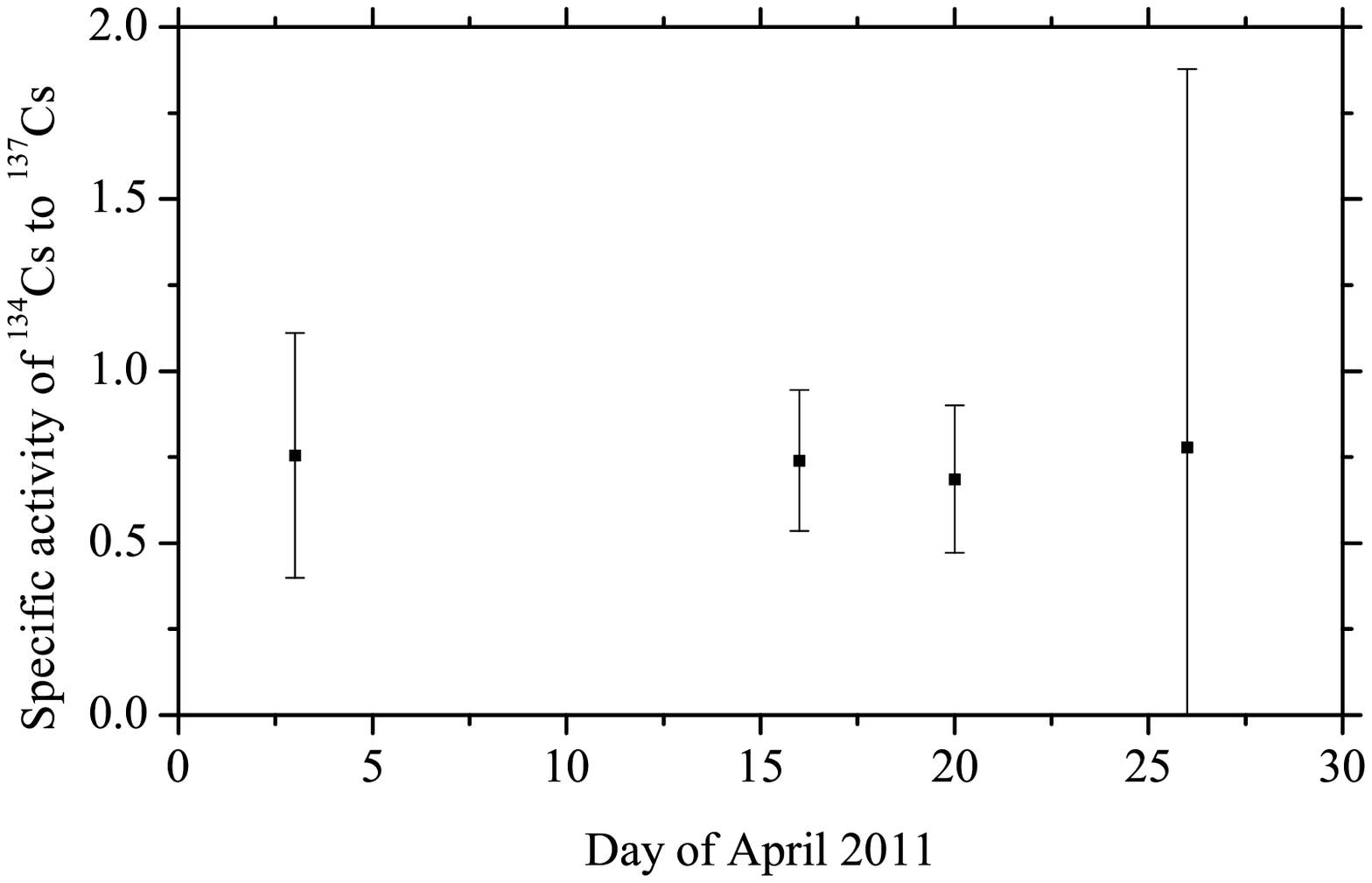}

\caption{Ratio of specific activity of \nuc{134}{Cs} to \nuc{137}{Cs}
vs time.}

\label{134to137_vs_time}
\end{figure}

The ratio of activity of \nuc{134}{Cs} to \nuc{137}{Cs} is shown in
Figure~\ref{134to137_vs_time}.  The weighted average ratio for the
four measurements is $0.72 \pm 0.13$, in agreement with the value of
$\sim$0.7 reported in \cite{Leon} and the measurements given in
\cite{Momoshima}.  This ratio is approximately constant because both
of these isotopes are products of nuclear fission, \nuc{137}{Cs}
directly, and \nuc{134}{Cs} by fission production of \nuc{133}{Cs}
followed by neutron capture.

Some laboratories \cite{Leon,Norman,MacMullin} have detected
\nuc{132}{Te} from the Fukushima incident, but we did not observe
this isotope. Our supposition is that this is because of its short 
half-life of 3.2~d, the appreciable delay between the release and 
our first measurements, and the low volatility of Te.

At no time during our measurements were the activities of the
isotopes we detected from Fukushima of any radioactive concern to the
inhabitants of northern Ontario.  The average radioactivity levels
were much less than what is received by normal background radiation
and we were only able to observe these isotopes because of our
extremely sensitive well-shielded low-background apparatus.

%At no time during our measurements were the activities of the
%isotopes we detected from Fukushima of any radioactive concern to the
%inhabitants of northern Ontario. The radioactivity levels were much
%less than what is received by normal background radiation and we were
%only able to observe these isotopes because of our extremely
%sensitive well-shielded low-background apparatus.

\section{Summary}

Several nuclear reactor fission products were observed in rainwater
samples collected in Greater Sudbury and $\gamma$-ray coun\-t\-ed in a
high-purity germanium detector at SNOLAB.  The short-lived isotope
\nuc{131}{I} and the longer-lived isotopes \nuc{134}{Cs} and
\nuc{137}{Cs}, were detected at concentrations much higher than in a
background sample.  The presence of all these isotopes is associated
with their release to the atmosphere from the nuclear accident at the
Fukushima Dai-ichi reactors in Japan.  These data, along with
measurements made in other places around the world, may aid our
understanding of the release of radioactive fission products and
their transport in the atmosphere.

\section*{Acknowledgments}

This work utilises infrastructure supported by the Natural Sciences and Engineering
Research Council, the Ontario Ministry of Research and Innovation,
the Northern Ontario Heritage Fund, and the Canada Foundation for
Innovation.  We thank the SNOLAB technical staff for developing the
infrastructure and their aid in our scientific endeavors, and Vale
S.~A. for hosting SNOLAB.

%\section*{References}

\bibliographystyle{elsarticle-harv}
%\bibliography{fukushima-v6}

\end{document}